\documentclass[twocolumn,showpacs,preprintnumbers,amsmath,amssymb]{revtex4}
\usepackage{graphicx}
\usepackage{epstopdf}
\usepackage{bm}

\begin{document}

\title{Contribution 6.4.2\\
Rydberg excitation of a Bose-Einstein condensate}

 \author{M. Viteau$^1$}
 
\author{M. Bason$^1$}
   
\author{J. Radogostowicz$^{2,3}$}
 
\author{N. Malossi$^{1,2}$}

 \author{O. Morsch$^1$}

\author{D. Ciampini$^{1,2,3}$}

\author{E. Arimondo$^{1,2,3}$}
 \email{arimondo@df.unipi.it}
 
 \affiliation{$^1$CNR-INO, Dipartimento di Fisica E.Fermi, Universit\`{a} di 
Pisa, Lgo Pontecorvo 3, I-56127 Pisa,Italy}

\affiliation{$^2$CNISM, Unit\'{a} di Pisa, Dipartimento di Fisica E.Fermi Largo Pontecorvo 3, I-56127 Pisa, Italy}

 \affiliation{$^3$Dipartimento di Fisica E.Fermi, Universit\`{a} di 
Pisa, Largo Pontecorvo 3, I-56127 Pisa, Italy}

\begin{abstract}
We have performed two-photon excitation  via the 6$^2$P$3/2$ state to $n$=50-80 S or D Rydberg state in Bose-Einstein condensates of rubidium atoms.   The Rydberg excitation  was performed in a quartz cell, where  electric fields generated by  plates external to the cell created electric charges on the cell walls. Avoiding accumulation of the charges and realizing good control over the applied electric field was obtained when the fields were applied only for a short time, typically a few microseconds.  Rydberg excitations of the Bose-Einstein condensates  loaded into quasi one-dimensional traps and in optical lattices have been investigated. The results for condensates expanded to different sizes in the one-dimensional trap agree well with the intuitive picture of a chain of Rydberg excitations controlled by the dipole-dipole interaction.  The optical lattice applied along the one-dimensional geometry produces localized, collective Rydberg excitations controlled by the nearest-neighbour blockade.
\end{abstract}


\date{\today}

\maketitle
\section{Introduction}
In recent years, Rydberg atoms have been the subject of intense study~\cite{Gallagher1994,GallagherPillet2008} and have
become an important tool for several branches of quantum physics such as quantum-information processing with neutral atoms~\cite{Jaksch2000,Lukin2001,Saffman2010},  production of single-atom as well as single-photon sources~\cite{Lukin2001,SaffmanWalker2002,PohlLukin2010},  quantum simulation of complex spin systems~\cite{WeimerBuechler2010}, creation of large optical nonlinearities~\cite{Mohapatra2008}, creation of entangled many-particle states~\cite{OlmosLesanovskyPRL2009} and measurements of electric fields close to the surface~\cite{Tauschinsky2010}.\\
\indent  The central ingredient of these schemes is the dipole blockade, i.e., when the excitation of an atom to a Rydberg state is inhibited if another, already excited, atom is less than the so-called blockade radius, $r_b$, away \cite{ComparatPillet2010}.  \indent  In the case of a large number of atoms, a simple physical picture is that of a collection of super-atoms \cite{vuletic_2006} in which a single excitation is shared among all the atoms inside the blockade radius \cite{robicheaux_2005,stanojevic_2009}. The Rydberg excitation blockade was realized in disordered clouds of cold atoms~\cite{Tong2004,Singer2004,Afrousheh2004,CubelLiebisch2005,Vogt2006,Heidemann2007,vanDitzhuijzen2008,Reinhard2008}  
as well as in a Bose Einstein condensate~\cite{Heidemann2008}. In addition the excitation of two individual atoms in the Rydberg blockade regime was recently performed~\cite{Urban2009,Gaetan2009} and their entanglement  demonstrated~\cite{Wilk2010,Isenhower2010}.\\
\indent The scalability of a quantum computer based  on the long-range interactions  of Rydberg atoms will rely on the realization of an array of trapped atoms~\cite{Saffman2010}. An optical lattice loaded with one atom in each site, such as the Mott insulator phase of ultracold bosonic atoms, represents the realization such an array.  Additional crucial issues to be solved in order to realize this quantum computation configuration are the initialization and readout operations on a single site of the lattice. However Lukin et al ~\cite{lukin_2001} suggested that experiments requiring single atoms excited to Rydberg states can also be performed with such collective Rydberg excitations, each containing several hundreds or thousand of atoms. This makes it possible to work with, e.g., Bose condensates loaded into optical lattices without the need to induce a Mott insulator transition in order to have single-atom occupation of the lattice sites. \\
\indent The present work demonstrates the realization of one-dimensional chains of collective Rydberg excitations in rubidium Bose condensates. In addition, the excitation dynamics of up to 50 Rydberg excitations in a condensate occupying around 100 sites of a one-dimensional optical lattice is studied. The excitation number is determined by the ratio between blockade radius and lattice spacing.\\
\indent  Our experiments are dependent upon the ability to combine techniques for obtaining Bose-Einstein condensates in dipole traps and optical lattices with those for creating and detecting Rydberg atoms. While the former require a vacuum apparatus with large optical access, the latter demand precisely controlled electric fields. As described in Sec. II we reconciled these requirements by adding external field electrodes to the quartz cell of our cold atom apparatus and carefully studying the operating conditions under which the best field control and highest detection efficiency of the Rydberg atoms was achieved. Sec. III reports the atomic parameters of the Rydberg states investigated in the present work, and also discusses the dependence of the laser excitation on the Zeeman level distribution.  Sec. IV presents experimental results for  Rydberg excitation in different BEC configurations and our results for the dipole blockade.  Conclusions complete the present work.
\section{Apparatus}
\begin{figure}[ht]
\centering\begin{center}
\includegraphics[width=8cm]{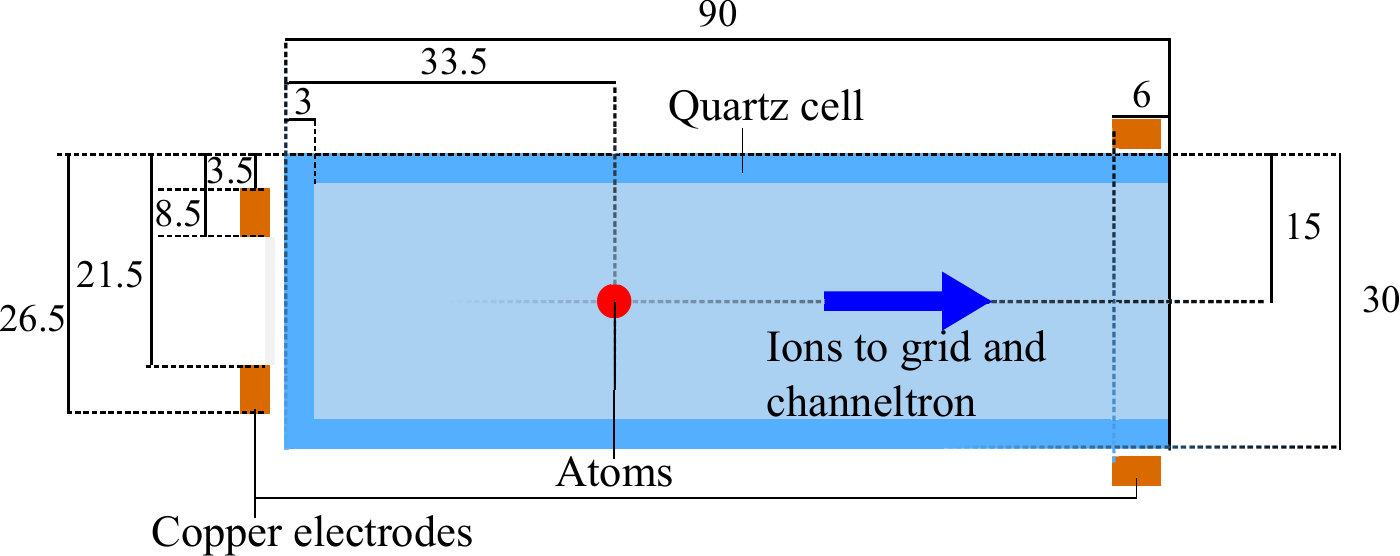}
\end{center}
\caption{Schematic of the vacuum quartz and collection system for the ion produced by the BEC excitation to Rydberg states, with dimensions in mm. Notice the copper plates  on the front  and on the sides of the cell. The grid and  the channeltron collecting the ions are located respectively 10 cm and 15 cm from the atomic cloud.  The laser beams are combined in order to excite rubidium atoms and to produce optical lattice within the same volume inside the cell.}
\label{ChargeCollection}
\end{figure}

 The  MOT/BEC apparatus uses two quartz cells, named collection and science respectively, horizontally
mounted on a central metallic structure~\cite{Lignier2007,Viteau2010}. We first create Bose-Einstein condensates of up to $10^5$ 87-Rb atoms using a two-step evaporation protocol with a TOP-trap and a crossed optical dipole trap. Once condensation is reached, the power in one of the trap beams is ramped down in $40\,\mathrm{ms}$ (in order to avoid excitation) and then switched off completely, while the power in the other beam is ramped up in order to keep the radial trap frequency almost constant. The condensate is then allowed to expand for variable times (up to $500\,\mathrm{ms}$) inside the (now one-dimensional) dipole trap, whose horizontal alignment has been optimized in order to avoid a centre-of-mass motion of the condensate along the trap direction.\\
\indent Rydberg states with principal quantum number $n$ between $55$ and $80$ are excited using a two-colour coherent excitation scheme with one blue laser at $421\,\mathrm{nm}$,  detuned by $0.5-1\,\mathrm{GHz}$ from the $6S_\mathrm{1/2} \to 6P_\mathrm{3/2}$ transition, and a second infrared (IR) laser at $1010-1030\,\mathrm{nm}$. The $421\,\mathrm{nm}$  blue radiation is generated by doubling a MOPA laser (TOPTICA TA 100, with output power 700 mW) with a TOPTICA cavity (output power 60 mW). The IR laser  allows ionization or atomic excitation to states with quantum number $n$ between 30 and the continuum.  High power and frequency stability of the IR radiation is achieved by injection locking a diode on an external cavity (output power 40 mW)  into a Sacher TIGER laser (output power 250 mW) where the grating is replaced by a mirror.  Both lasers systems, having a line-width smaller that 1 MHz, were locked using a Fabry Perot interferometer where a 780 nm laser locked to the Rb resonance acted as a reference.\\
\indent Both blue and IR excitation beams were aligned  so as to be almost parallel to the dipole trap beam in which the condensate expanded, and focused to a waist  larger than the atomic cloud size, the minimum being 100 $\mathrm{\mu{m}}$ in order to maintain the  ultracold cloud within the beam waist in  presence of fluctuations in the spatial position of the cold atomic cloud.  The blue/IR laser beams were pulsed for durations in the range of 0.1-10 $\mu$s with a gaussian rise time of 0.08 $\mu$s controlled by an acousto-optic modulator. The Rabi frequencies for the blue and IR lasers are $\Omega_ \mathrm{blue}/2\pi$ up to 80 MHz and $\Omega_ \mathrm{IR}/2\pi$ around 10 MHz, respectively.  For a blue laser detuning of 1 GHz, the two-photon Rabi frequency is in the 20-100 kHz range.
The polarizations of both blue and IR lasers were linear and parallel to each other along the vertical axis. The excitation of the condensate to Rydberg states was performed 0.7 ms after the TOP magnetic field was switched off, with the atomic magnetic moments rotating in a plane containing the laser polarizazion axis.\\  
\indent The Rydberg atoms are ionized and then detected by applying, just after the Rydberg excitation
pulse, an electric field to the cell plates shown in Fig.~\ref{ChargeCollection}. In order to avoid the formation of electric charges on the cell walls, which would have led to the creation of uncontrolled electric fields, the Rydberg field ionization was based on an electric field pulse with a rise time of 700 ns and a few $\mu$s duration applied to the copper plates outside the cell. In previous work~\cite{Viteau2010}, we measured a high collection efficiency $\eta=0.35(10)$ which includes both the fraction of the produced charges that are transported to the channel electron multiplier (CEM) through a two-stage acceleration and the multiplier detection efficiency. \\

 \indent The absorption of a  421 nm photon by an atom in the $\mathrm{6^2P}_{3/2}$ state ionizes the rubidium atoms with an  ionization cross-section of 4.7 Mbarn~\cite{Courtade2004},  while the cross-section for absorption of  1 $\mu$m photon from the $n\approx 50$ Rydberg states is very small,  $4.5 \times 10^{-5}$ Mbarn for 50$^2$S and
$1.2 \times 10^{-2}$ Mbarn for 50$^2$D~\cite{Saffman2005}.  We calculated  the average number of ions produced by the two-step photoionization pulse from the solution of the master equations describing the atomic interaction with the laser pulses introduced in~\cite{Courtade2004}.  As an example the ionization probability for a 1 $\mu$s blue laser pulse  is only $1.\times10^{-7}$  for a Rydberg excitation with $\Omega_{\rm blue}/2\pi=10$ MHz and 0.5 GHz detuning from the 6$^2$P$_{3/2}$ intermediate state.  For longer duration laser pulses the Penning ionization caused by the interaction-induced motion in cold Rydberg gases is an additional ionization process~\cite{LiGallagher2005}.  \\
\indent Rydberg states are very sensitive to the presence of electric fields that produce either a broadening or a splitting of the excitation lines and therefore, at given laser parameters, a modification of the excitation efficiency. During the Rydberg excitation we have applied to the external plates low voltages, 10 Volts maximum, and monitored the frequency shift of the Rydberg spectra.  The Stark shifts of the absorption lines provide  evidence to show that our resolution for the applied electric field is at the 5 mV/cm level. Ions generated by laser photoionization of rubidium  atoms  could create an electric field modifying the absorption of  the remaining atoms, but owing to their negligible number in our experimental conditions their role is neglected.\\
\indent We create one-dimensional lattices by intersecting two linearly polarized laser beams of wavelength $\lambda_L=840\,\mathrm{nm}$ at an angle $\theta$, leading to a lattice with spacing $d_L=\lambda_L/(2\sin(\theta/2))$. The optical access in our setup allows us to realize spacings up to $\approx 25\,\mathrm{\mu m}$. For the largest spacing, the depth $V_0$ of the periodic potential $V(x)=V_0\sin(2\pi x/d_L)$ measured in units of the recoil energy $E_\mathrm{rec}=\hbar^2\pi^2/(2m d_L^2)$ (where $m$ is the mass of the atoms) is up to $V_0/E_\mathrm{rec}\sim 15000$. In the lattice direction the size $\Delta x$ of the on-site wavefunction is around $\Delta x = d_L/(\pi(V_0/E_\mathrm{rec})^{1/4})\sim 0.05-0.4\,\mathrm{\mu m}$. \\

\section{Atom-laser interactions}
\label{interaction}
 \begin{table}
\caption{Parameters of the explored Rydberg states}
\centering\begin{center}
\begin{tabular}{ccccccc}
\hline\noalign{\smallskip}
 Transition &$\Gamma_{0}$&$\Gamma_{\rm BBR}$&$A(nL \to 6^2P_{3/2})$&Dipole Moment \\
 6$^2$P$_{3/2}$ $\to nL$&(s$^{-1}$)&(s$^{-1}$)&(s$^{-1}$) &(e a$_0$ units) \\
\noalign{\smallskip}\hline\noalign{\smallskip}
53D$_{3/2}$&7.02$\times$10$^3$&5.60$\times$10$^3$&4.78$\times$10$^2$&3.51$\times$10$^{-2}$\\
53D$_{5/2}$&7.07$\times$10$^3$&5.40$\times$10$^3$&4.73$\times$10$^2$&3.49$\times$10$^{-2}$\\
78D$_{3/2}$&2.16$\times$10$^3$&3.46$\times$10$^3$&1.28$\times$10$^2$&1.81$\times$10$^{-2}$\\
78D$_{5/2}$&2.17$\times$10$^3$&3.46$\times$10$^3$&1.28$\times$10$^2$&1.81$\times$10$^{-2}$\\
\noalign{\smallskip}\hline
\end{tabular}
\end{center}
\label{table}
\end{table}

The effective decay rate $\Gamma_{nL}$   of the $nL$ Rydberg level  is  given by the sum of the $\Gamma_0$, the  total spontaneous emission rate, and $\Gamma_{\rm BBR}$, the depopulation rate  induced by black body radiation (BBR)~\cite{Beterov2009}
\begin{equation}
\Gamma_{nL} =\Gamma_0 +\Gamma_{\rm BBR}.
\end{equation}
$\Gamma_0$ is determined by the $A(nL \to n^\prime L^\prime)$ Einstein coefficients  from the $|nL>$  level to all lower-lying levels
\begin{equation}
\Gamma_0 =\sum_{ n^\prime L^\prime \neq nL}A(nL \to n^\prime L^\prime).
\end{equation}
$\Gamma_{\rm BBR}$ at the temperature $T$ is written in a similar form, including both lower and higher states
\begin{equation}
\Gamma_{\rm BBR} =\sum_{ n^\prime L^\prime \neq nL}A(nL \to n^\prime L^\prime)\frac{1}{exp\left(\frac{E_{nL}-E_{n^\prime L^\prime}}{kT}\right)-1}%\frac{E_{\rm nL}-E_{\rm n^\prime L^\prime}}{kT}-1}.
\end{equation}
 The $A(nL \to 6^2P_{3/2})$ spontaneous decay rate from the Rydberg level to all the Zeeman states of the 6$^2$P$ _{3/2}$  level determines the transition dipole moment, appearing in the definition of  the Rabi frequencies. The atomic parameters for the few Rydberg states investigated are listed in Table I.\\
\begin{figure}
\centering
\includegraphics[width=8cm]{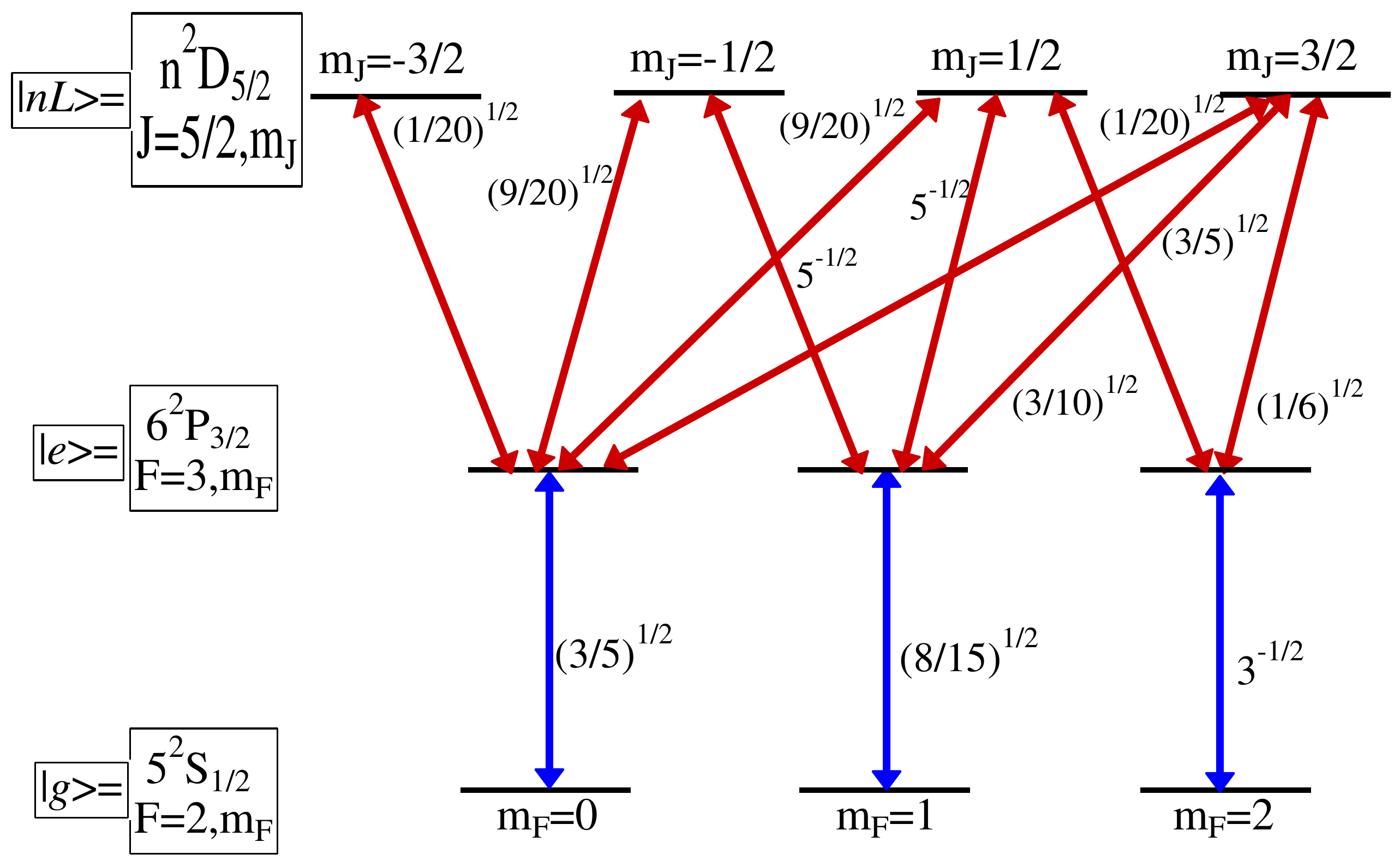}
\caption{Atomic states, selection rules imposed by the laser polarization and Clebsch-Gordan coefficients for 5$^2$S$_{1/2}(F=2) \to$ 5$^2$P$_{3/2}(F=3)\rangle \to n^2$D$_{5/2}$ two-photon excitation, by supposing the blue and IR lasers linearly polarized along the same direction in space and an arbitrary initial occupation of Zeeman levels.  The $|F,m_F>$ label shown in the bottom is applied to denote the $^2$S and $^2$P states;  the $|J,m_J>$ label shown in the top applies  to the $^2$D states. }
\label{ExcitationDipoles}
\end{figure}
 \indent The Rabi frequencies of the  $|g\rangle \to |e\rangle \to |nL\rangle$ transitions depend on  the quantum numbers of the atomic states composing the excitation sequence. For the
 $|g\rangle=$5$^2$S$_{1/2}(F=2)$ and $|e\rangle=$6$^2$P$_{3/2}(F'=3)$  states the quantum numbers are the  hyperfine ones $|F,m_F\rangle$. Instead for the Rydberg states where the hyperfine splitting is small, the states are characterized by the   $J,m_J$ quantum numbers.  As a consequence the blue and IR Rabi frequencies  should be calculated in two different bases, $|I,J,F,m_F\rangle$ to be denoted by the $F,m_F$ subscript and $|I,m_I,J,m_J\rangle$ to be denoted by  the $I,J$ subscript. Therefore in order to derive the atomic coupling with the IR laser, the intermediate 6$^2$P$_{3/2}(F=3, m_F)$  state excited by the blue laser  should be decomposed into the states of the $|I,m_I,J,m_J\rangle$ basis.
 As an example, we report in Fig.~\ref{ExcitationDipoles} the laser excitation for the $|g\rangle \to |e\rangle \to |n^2D_{5/2,3/2}\rangle$ transitions in the case of blue and IR laser fields with linear polarization for excitation from the different Zeeman levels of the ground state. 
The Clebsch-Gordan coefficients associated with the excitation paths are marked in the figure. The dipole moments determining the $\Omega_{\rm bl}$ and $\Omega_{\rm IR}$ Rabi frequencies are obtained by multiplying the atomic dipole moment by the Clebsch-Gordan coefficient.  The Rydberg excitation path is determined   by the orientation of the laser electric field with respect to the atomic orientation. \\

\begin{figure}[ht]
\centering\begin{center}
\includegraphics[width=8.5cm]{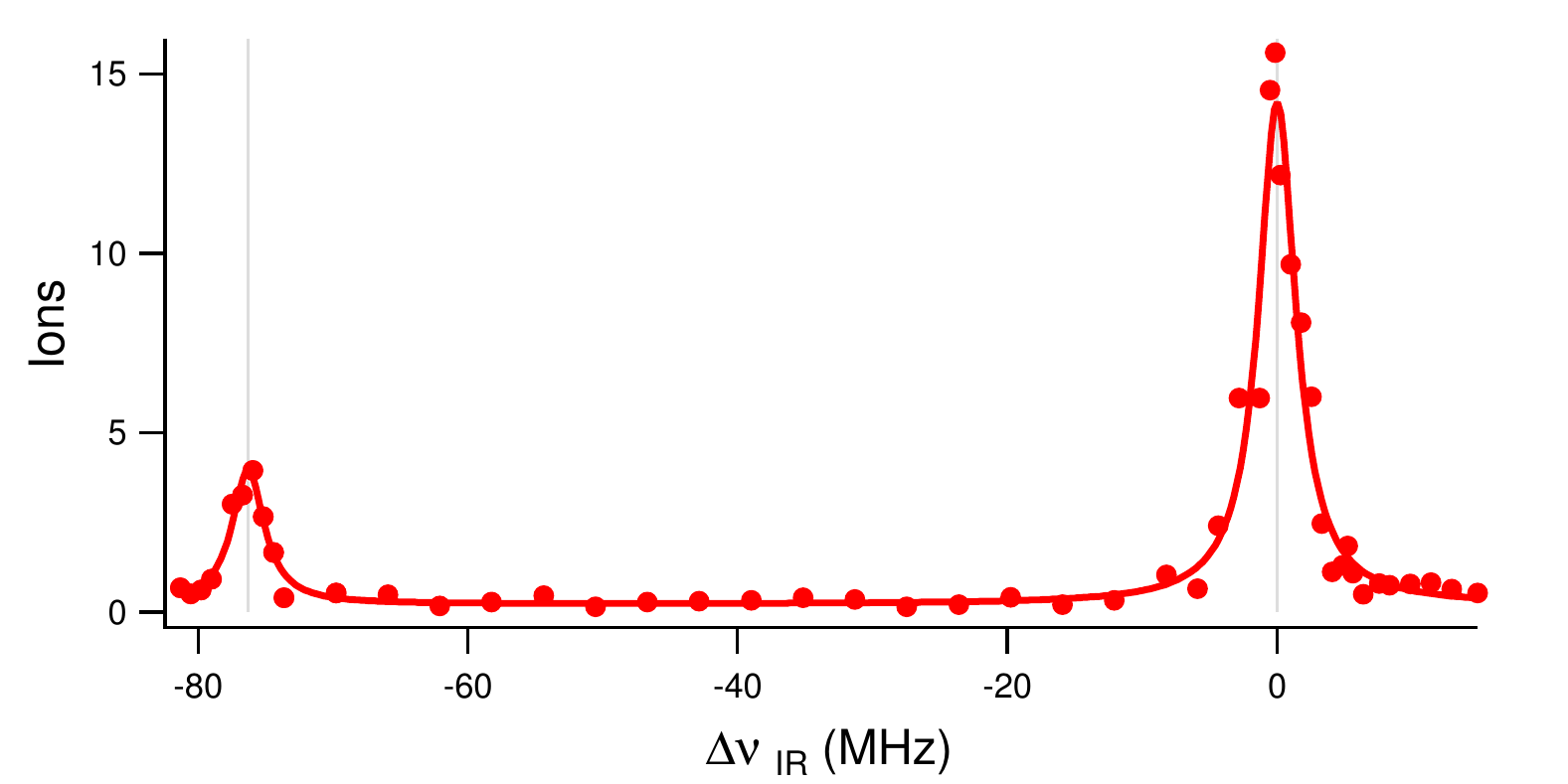}
\end{center}
\caption{Detected  ion number versus   the $\delta_{\rm IR}$ IR laser detuning produced by the  excitation to the 53$^2$D$_{5/2}$ Rydberg state, right peak, and to the 53$^2$D$_{3/2}$ Rydberg state, left peak. The absolute ion number is determined with a fifteen percent accuracy.  Parameters  $\delta_{\rm blue}=500$ MHz, $\Omega_{\rm blue}= 40 $ MHz and $\Omega_{\rm IR}= 6.7 $ and 3.5 MHz for the $D_{5/2}$ and $D_{3/2}$ excitations, respectively.}
\label{53DSpectrum}
\end{figure}

\section{Rydberg excitation}
\subsection{Rydberg spectra}
We investigated the range of Rydberg energy levels we could tune the IR laser to. For this purpose, following excitation to the Rydberg state for the atoms in the BEC, we measured the ions produced by field ionization, with negligible contributions from 6P photoionization,  blackbody radiation absorption and Rydberg Penning ionization.  Typically we acquired Rydberg spectra by scanning the frequency of the IR laser at fixed frequency of the blue laser. Fig. \ref{53DSpectrum}  reports the spectrum of the detected ions as a function of the IR laser frequency scanned around the the resonant frequency for the Rydberg excitation from the 6$^2$P$_{3/2} (F=3)$ state to the 53$^2$D$_{3/2,5/2}$ and 53$^2$D$_{5/2}$ states. The spectrum was obtained following a 100 ms expansion of the condensate.  The fine splitting between the 53D$_{3/2,5/2}$  levels is in agreement with the predictions of Li {\it et al.}~\cite{LiGallagher2003}. Spectra recorded within the condensate without expansion reported ion production also for laser excitation on the red side of both fine structure peaks.   \\

\begin{figure}[htp]
\centering\begin{center}
\includegraphics[width=8.5cm]{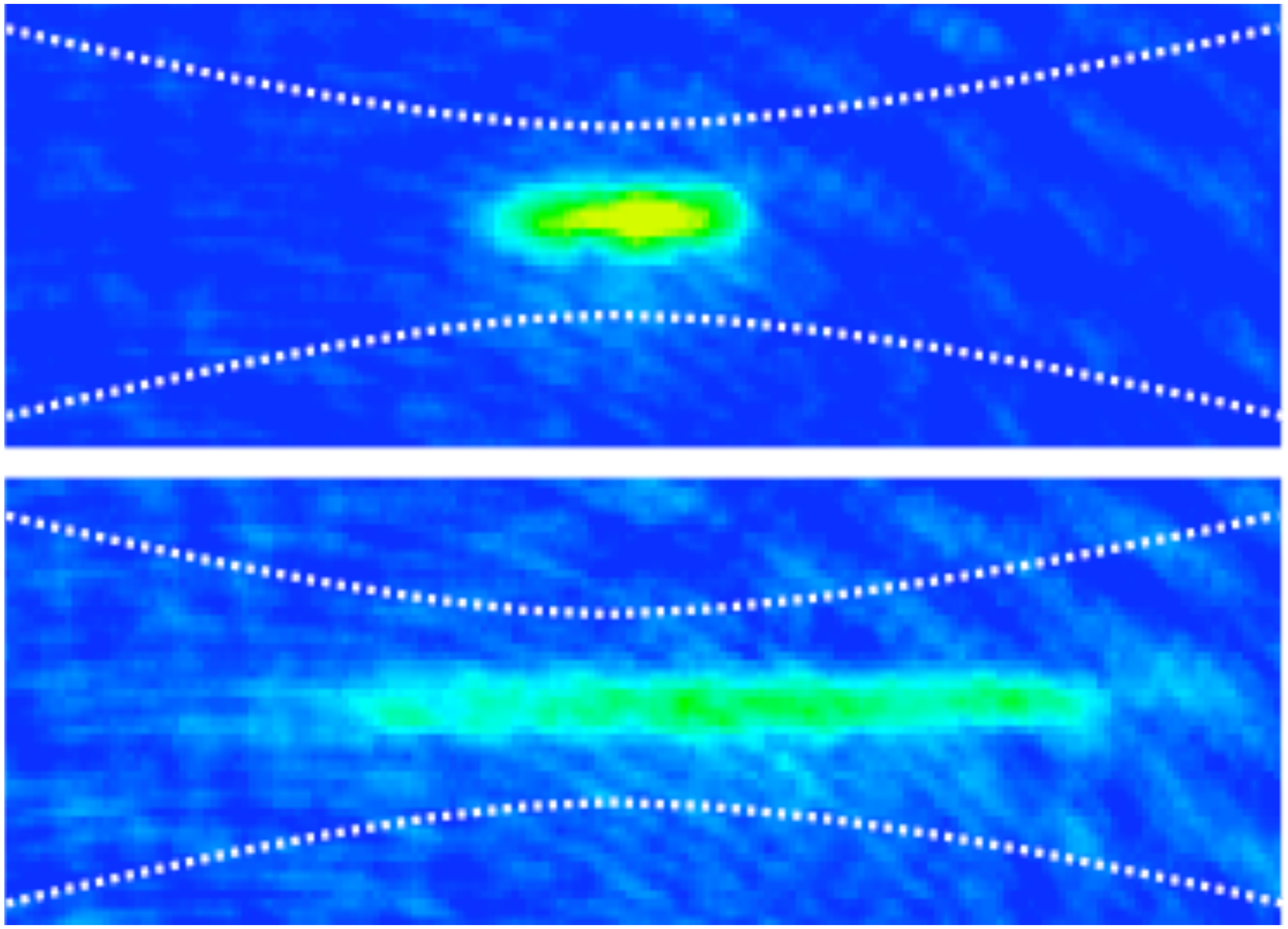}
\end{center}
\caption{Images of the condensate following different 1D expansion times, 10 ms  in the top  and 50 ms in the bottom. The condensate size appearing in the image is limited by the spatial resolution of the imaging system. The dashed white lines represent schematically the transverse trap confinement. Radially the size of the condensate is smaller than the blockade radius $r_b$, leading to an effectively one-dimensional array of Rydberg excitations. As the condensate expands, more collective Rydberg excitations fit into the condensate length.}
\label{Images}
\end{figure}

\subsection{1D geometry}
 Quasi one-dimensional atomic samples of length $l$ are created by switching off one of the two trap beams of the crossed dipole trap in which the condensates are created and letting the condensate expand for variable times (up to $500\,\mathrm{ms}$), as shown in Fig. \ref{Images}. The resulting highly elongated atomic clouds are up to $l=1\,\mathrm{mm}$ long, while their radial dimensions are on the order of $1-2\,\mathrm{\mu m}$ (radial dipole trap frequencies are around $100\,\mathrm{Hz}$).\\
 \indent  Since the expected blockade radii for the Rydberg states between $n=50$ and $n=80$ used in our experiments range from $5\,\mathrm{\mu m}$ to $15\,\mathrm{\mu m}$, this suggests that at most one Rydberg excitation ``fits'' radially into the condensate, and several Rydberg excitations inside such a sample organize themselves in a linear array. Notice that the condensate spatial distribution is not in the 1D regime, and the 1D character refers only to the Rydberg excitation.\\
 
\subsection{Coherent excitation}
\begin{figure}
\centering
\includegraphics[width=8.5cm]{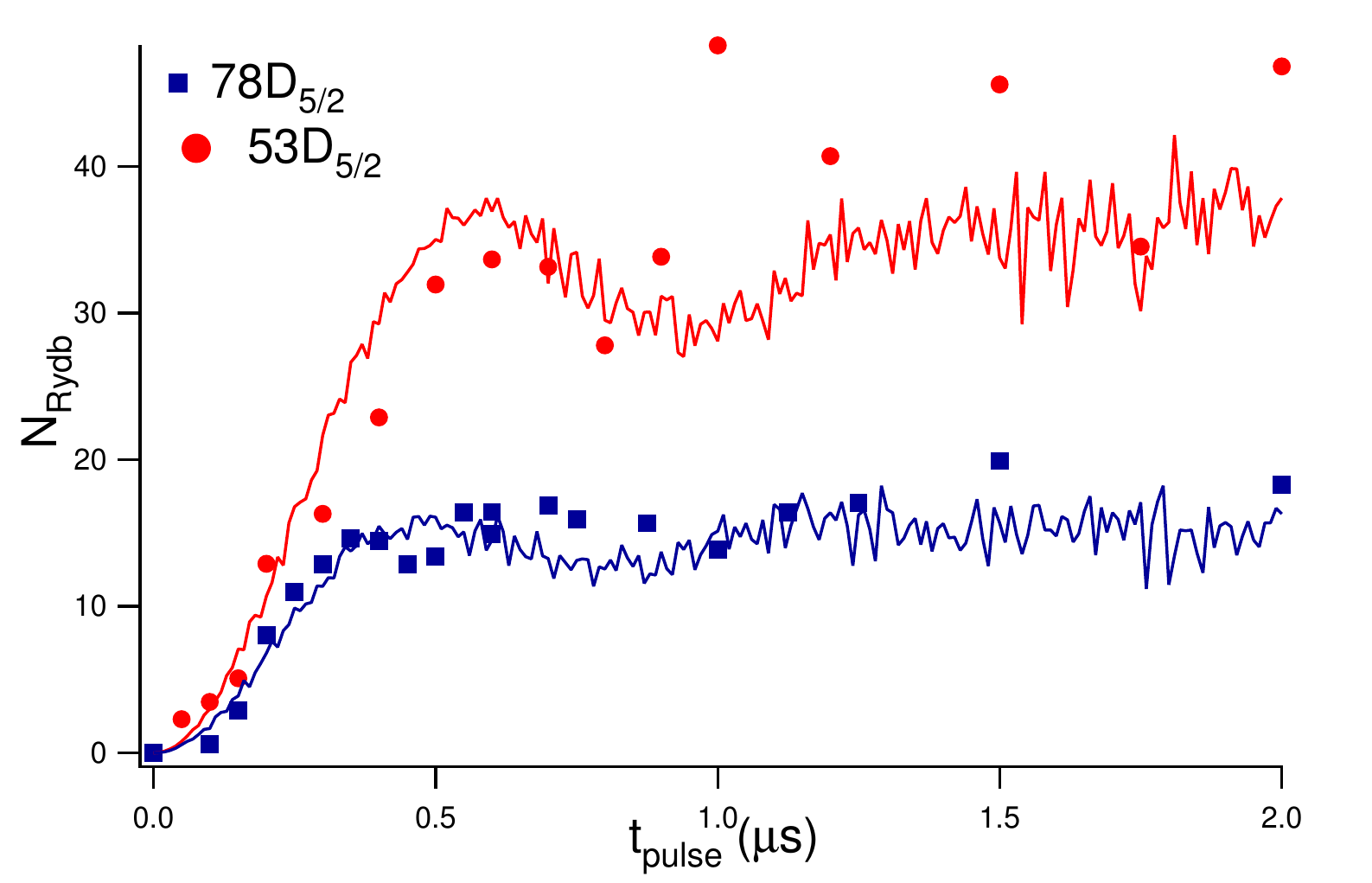}
\caption{ The number of Rydbergs $N_{\rm Rydb}$ vs the pulsed excitation length, $t_{\rm pulse}$, for the 53D$_{5/2}$ (red dots) and 78D$_{5/2}$ Rydberg states (blue squares). $N_{\rm Rydb}$ was derived by scaling the detected ions by the collection efficiency. The BEC length  was $\sim$~160~$\mu$m after a 50~ms expansion. The  results of theoretical simulations are represented by the continuous lines.}
\label{CoherentExcitation}
\end{figure}

After expanding the condensates, Rydberg atoms are created using pulses between 100~ns and 3~$\mu$s duration (during which the condensate expansion is frozen and Penning and blackbody ionization \cite{WalkerSaffman2008} were found to be negligible) and finally detected by field ionization. The duration of the excitation pulses is chosen such that the system is in the saturated regime in which the number of Rydberg atoms levels off after an initial increase on a timescale of hundreds of nanoseconds \cite{Heidemann2007}. \\
\indent The observation of coherent excitation to the Rydberg state is based on the variation of the irradiation pulse durations $t_{\rm pulse}$  at a given condensate expansion time, i.e.,  at a given atomic density.  Fig. \ref{CoherentExcitation} reports the observations of excitation dynamics for the 53D$_{5/2}$  and 78D$_{5/2}$ states, where after the initial increase, determined by the Rabi evolution between ground and Rydberg states, the number of produced Rydberg atoms experiences a slow increase with the excitation time.  According to the super-atom picture the collective Rabi frequency for the coherent  excitation of $N$ atoms is
\begin{equation}
\Omega_\mathrm{coll}=\sqrt{N}\Omega_\mathrm{single},
\end{equation}
 where the single-particle Rabi frequency $\Omega_\mathrm{single}\sim 2\pi\times 200\,\mathrm{kHz}$ for our experimental parameters.  From the experimental points one can see that the Rabi oscillations that should characterize the coherent excitation are not visible due to the non uniform distribution of the condensate atoms in the sample. Thus the structures presented in Fig. \ref{CoherentExcitation} are the average of excitations with different collective frequency and the Rabi oscillations are washed out.  As the blockade radius for the 53D$_{5/2}$ is half of the blockade radius of the 78D$_{5/2}$ state, the number of blocked regions that fits into the sample volume is twice as large, therefore we observe twice as many detected Rydberg atoms. This is confirmed by a simple theoretical model assuming complete collective Rabi oscillations between the ground and excited state without decoherence and including an average over the  spatial dependence of the collective Rabi frequency. The Rabi frequency was slightly adjusted due to the imprecise measurement of the waist and powers of both excitation laser.

\begin{figure}[htp]
\centering\begin{center}
\includegraphics[width=8.5cm]{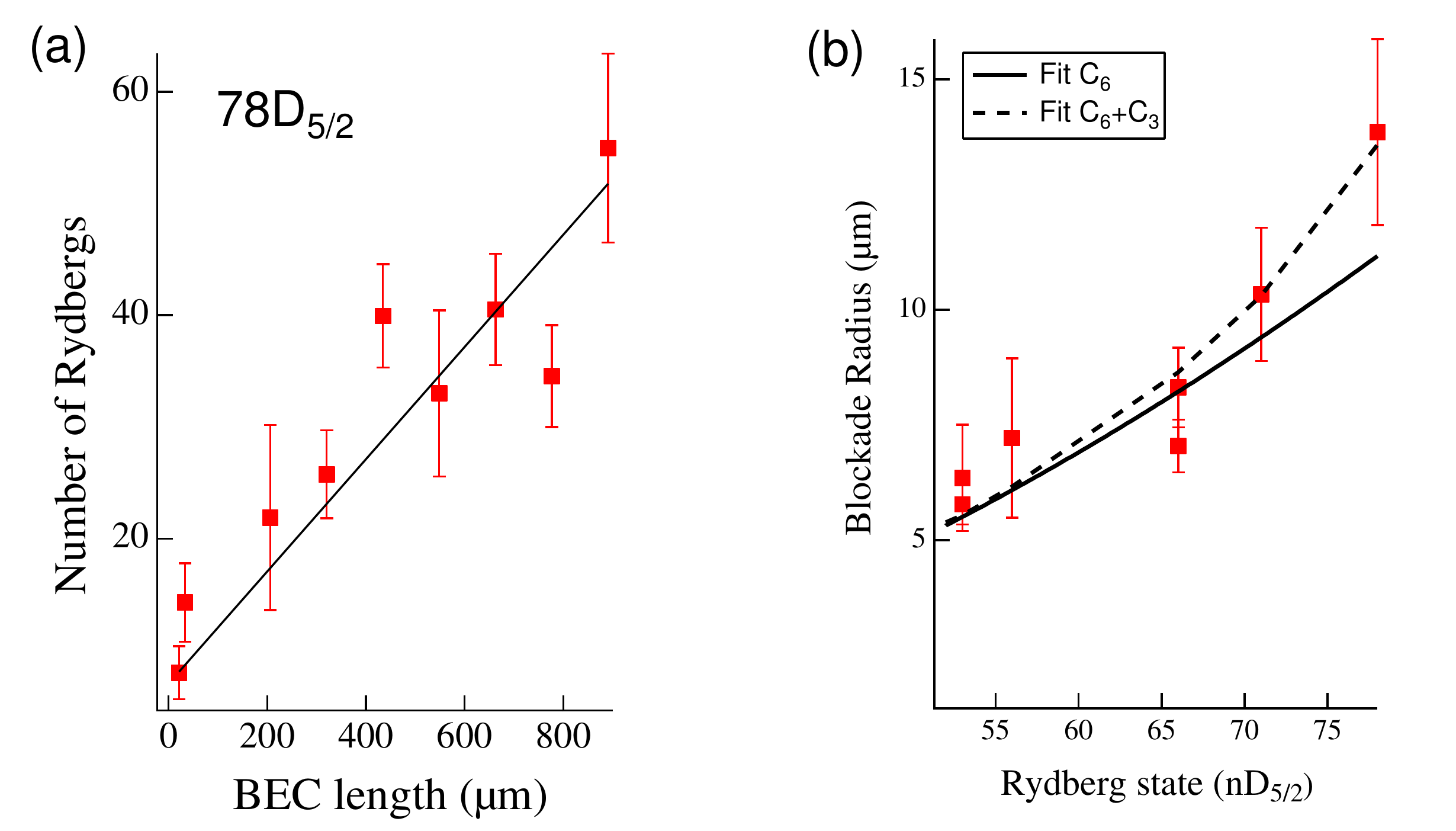}
\end{center}
\caption{In (a), the number of Rydberg excitations vs the condensate length. In (b), the measured blockade radius $r_b$ vs the principal quantum number $n$. The continuous line is the theoretically predicted value assuming a laser linewidth of $300\,\mathrm{kHz}$. The dashed line is the theoretically predicted value assuming the presence of an additional dipole-dipole interaction produced by a stray electric field of 5~mV/cm at the limit of our precision.} 
\label{DipoleBlockade}
\end{figure}

 \subsection{Dipole blockade}
Fig. \ref{DipoleBlockade}(a) shows the results of an excitation experiment for the $66D_{5/2}$ state using a $1\,\mathrm{\mu s}$ excitation pulse, in which a linear increase of the number of Rydberg atoms with the length of the condensate is visible. This result agrees with the simple intuitive picture of super-atoms, with dimension twice the $r_b$ blockade radius, being stacked in a one-dimensional array of varying length.\\
 \indent  The highly correlated character of the Rydberg excitations thus created is further demonstrated by analyzing the counting statistics of our experiments, which are strongly sub-Poissonian for short expansion times of a few milliseconds with observed Mandel $Q$-factors \cite{Ates2006} of $Q_{obs}\approx -0.3$, where $Q_{obs}=\Delta N/\langle N\rangle-1$ with $\Delta N$ and $\langle N\rangle$ the standard deviation and mean of the counting statistics, respectively. Taking into account the detection efficiency, $\eta$, for Rydberg excitations in our system, this indicates actual $Q$-factors close to $-1$ (since the observed $Q$-factor $Q_\mathrm{obs}=\eta Q_\mathrm{act}$).\\
 \indent  From the data of  Fig. \ref{DipoleBlockade}(a) we can extract the mean distance between adjacent Rydberg atoms, where we have to take into consideration the fact that in the saturated regime the individual super-atoms perform Rabi oscillations between their ground and excited states and that hence on average only half of the excitable super-atoms are detected. Fig. \ref{DipoleBlockade}(b) shows the blockade radius measured in this way for different Rydberg states, together with the theoretically expected value \cite{ComparatPillet2010} for a pure van-der-Waals interaction following an $n^{11/6}$ scaling and assuming a linewidth of our excitation lasers of around $300\,\mathrm{kHz}$. The measured values for the blockade radii of $5-15\,\mathrm{\mu m}$ confirm our interpretation of an essentially one-dimensional chain of Rydberg atoms. For high-lying Rydberg states with $n>75$ we find deviations from the theoretically expected scaling with $n$, which may be due to small electric background fields that are difficult to avoid in our setup with the field electrodes located outside the glass cell in which the condensates are created.  In Fig. \ref{DipoleBlockade}(b) a theoretical analysis by supposing an applied weak electric field, around 5 mV/cm, produces a better fit of the experimental observations.
 By deliberately applying an electric field during the excitation pulse we have checked that for large $n$, even a small electric field on the order of 0.1 $\mathrm{V/cm}$ can lead to a dipole-dipole interaction energy contribution (scaling
as $n^{4/3}$) on the order of the usual van-der-Waals interaction, resulting in an effective blockade radius that is substantially larger than what is expected for the pure van-der-Waals case. 

\begin{figure}
 \centering
 \includegraphics[scale=0.4]{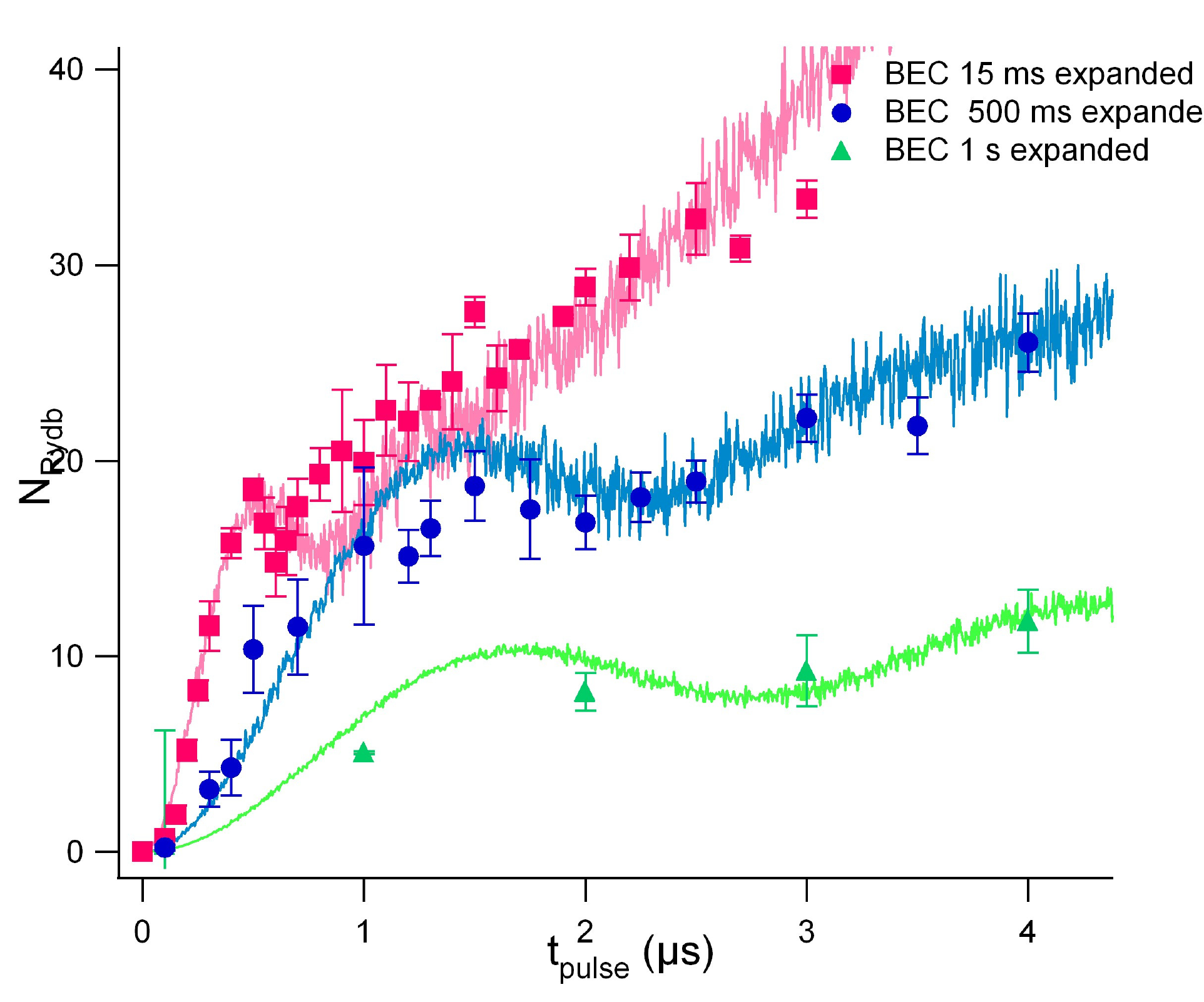}
 \caption[Level diagram of 87 Rb]{ The temporal dynamics of the Rydberg excitation for the Bose Einstein condensate loaded into an optical lattice. Atoms were excited by 421~nm laser light blue detuned by 1~GHz from the 6$^2$P$_{3/2}$ state.The light intensities were $I_{\rm blue}\approx$~40~W/cm$^{2}$ for the 421~nm light and $I_{IR}\approx$~380~W/cm$^{2}$ for the 1021~nm laser, leading to a two-photon Rabi-frequency $\Omega_{\rm single}/2 \pi=125 $ kHz. Results for various BEC expansions times of 15~ms (red squares), 500~ms (blue dots) and 1~s by (green triangles) are shown. The continuous lines represent the results obtained from theoretical simulations. }
\label{fig:6:10}
\end{figure}	

\subsection{1D optical lattice}  
Having demonstrated the one-dimensional character of the Rydberg excitation in the elongated condensate, now we add an optical lattice along the direction of the one-dimensional sample. The spacing of the lattice in our setup  can be varied between $d_L=0.42\,\mathrm{\mu m}$ and $\approx 25\,\mathrm{\mu m}$ and hence interpolates between the extremes $d_L\ll r_b$ in which the Rydberg excitation on one site interacts strongly with other excitation many sites away, and $d_L\gtrsim r_b$ where only nearest neighbours are expected to interact significantly. The depth of the optical lattice is sufficient to ensure that the ratio between the size $\Delta x$ of an on-site wavefunction and $d_L$ is less than $\approx 0.1$, meaning that the individual mini-condensates created by the lattice are well separated from one another and their longitudinal size is much less than $r_b$.\\
\indent  The effect of several Rydberg excitation cycles on the condensate phase coherence between the lattice sites was tested by performing a time-of-flight experiment after the excitations. We verified that more than $10$ excitation-detection cycles can be performed on the same condensate without a noticeable loss of phase coherence or atom number in the condensate. \\
\indent In Fig. \ref{fig:6:10} the experimental results of the excitation dynamics for three different BEC densities are presented along with the results obtained from a theoretical model, summarized below. The expansion of the condensates was chosen to be 15~ms, 500~ms and 1~s, respectively, leading to a decrease in  atom number by factors of 4 and 5 in each step.  
 The timescale of the collective excitation scales with $\sqrt{N}$, as expected, and the fraction of lattice sites containing a Rydberg atom is around $0.25$ in the saturated regime. In the transient regime we see the remnants of a coherent Rabi oscillation, which is washed out due to the different collective local Rabi frequencies. This effect stems from a less than uniform distribution of atom numbers. A simple numerical model containing a gaussian distribution of on-site Rabi frequencies and a small dephasing rate between the lattice sites, chosen such as to reproduce the observed long-term increase in the number of collective excitations, agrees well with our experimental observations. 
 
 \section{Conclusions}
We have demonstrated the controlled preparation of Rydberg excitations in large ensembles of ultracold atoms forming structures of localized collective excitations, either self-generated by the long-range interactions between Rydberg atoms or imposed by an optical lattice. Our results can straightforwardly be extended to two- and three-dimensional lattice geometries and to even larger lattice spacings that will allow selective Rydberg excitation on a single site. Furthermore, appropriate detection techniques such as microchannel plates should allow direct observation of the distribution of Rydberg excitations in the lattice.\\
\indent Classical and quantum correlations, and highly entangled  collective states are  expected to be created, as pointed out in~\cite{lesanovsky_2011} for one dimensional Rydberg gases and in~\cite{zuo_2010} for one-dimensional optical lattices. Our results pave the way towards their controlled creation.\\
 
\section{Acknowledgement}
Financial support by the EU-STREP ``NAMEQUAM'' and by
a CNISM ``Progetto Innesco 2007'' is gratefully acknowledged.

\end{document}